\documentclass[aps,prl,twocolumn,floats,amsmath,amssymb,superscriptaddress,reprint]{revtex4}
\usepackage{graphicx}
\usepackage{epsfig}
\usepackage{amsfonts}
\usepackage{amsmath}
\usepackage{natbib}
\usepackage{multirow}
\usepackage{bm}
\usepackage{epstopdf}
\usepackage{ulem}
\usepackage{pdfpages}

\newcommand{\exciting}{{\usefont{T1}{lmtt}{b}{n}exciting}}
\begin{document}
\title{Dimensionality of excitons in stacked van der Waals materials: \\ The example of hexagonal boron nitride}
\author{Wahib Aggoune}
\affiliation{Institut f\"{u}r Physik and IRIS Adlershof, Humboldt-Universit\"{a}t zu Berlin, Berlin, Germany}
\affiliation{Laboratoire de Physique Th\'eorique, Facult\'e des Sciences Exactes, Universit\'e de Bejaia, 06000 Bejaia, Algeria}
 
\author{Caterina Cocchi}
\author{Dmitrii Nabok}
\affiliation{Institut f\"{u}r Physik and IRIS Adlershof, Humboldt-Universit\"{a}t zu Berlin, Berlin, Germany}
\affiliation{European Theoretical Spectroscopic Facility (ETSF)}

\author{Karim Rezouali}
\author{Mohamed Akli Belkhir}
\affiliation{Laboratoire de Physique Th\'eorique, Facult\'e des Sciences Exactes, Universit\'e de Bejaia, 06000 Bejaia, Algeria}

\author{Claudia Draxl}
\affiliation{Institut f\"{u}r Physik and IRIS Adlershof, Humboldt-Universit\"{a}t zu Berlin, Berlin, Germany}
\affiliation{European Theoretical Spectroscopic Facility (ETSF)}
\begin{abstract}
With the example of hexagonal boron nitride, we demonstrate how the character of electron-hole ($e$-$h$) pairs in van der Waals bound low-dimensional systems is driven by layer stacking. Four types of excitons appear, with either a two- or three-dimensional spatial extension. Electron and hole distributions are either overlapping or exhibit a charge-transfer nature. We discuss under which structural and symmetry conditions they appear and they are either dark or bright. This analysis provides the key elements to identify, predict, and possibly tailor the character of $e$-$h$ pairs in van der Waals materials.
\end{abstract}
\maketitle


Two-dimensional (2D) systems and layered weakly-bound structures thereof are considered the materials of the 21st century. Their wealth of intriguing properties is widely explored from a fundamental scientific point of view but also in view of a plethora of possible applications~\cite{bhim+15nano,tan+17cr}. 
Hexagonal boron nitride (h-BN) is one of these materials, consisting of covalently bound sheets that are held together by van der Waals (vdW) forces~\cite{pease+50nat}.
h-BN is a wide-gap semiconductor, exhibiting pronounced excitonic effects in its optical excitations that are present irrespective of the material's dimensionality~\cite{wata+04natm,kubo+07sci,arna+06prl,wirtz+08prl,gala+11prb,bourrellier+14acs,fulvio+16prb,koskelo+17prb}. 
Owing to the flat geometry of its in-plane hexagonal lattice, h-BN is often chosen as a building block in vdW heterostructures~\cite{dean+10natn,yan+12prb,agg+17jpcl,lati+17nl}. 
Combining different 2D systems, quantum confinement effects allow for tailoring their opto-electronic properties~\cite{geim-grig13nat,novo+16sci,jari+17natm}. 
This not only concerns level alignment at the interface~\cite{kang+13apl,chiu+15natcom,ozcc+16prb,fu+16jpcc,guo-robe16apl,zhan+16tdm2,thyg17tdm,fu+17pccp} but also the way the system interacts with light, i.e., quantum efficiency, as well as the character and spatial distribution of electron-hole ($e$-$h$) pairs~\cite{milk+13jpcl,hong+14natn,ceba+14nano,li+16cm,bara+17nl,agg+17jpcl,guan+17jpcc,lati+17nl}.

In this Rapid Communication, we show that the nature and dimensionality of excitons can also be governed in a single vdW-bound bulk material, taking h-BN as an example. 
This seems surprising at a first glance as $e$-$h$ pairs in this material have been found to exhibit basically the same character and extension in bulk~\cite{arna+06prl,wirtz+08prl,gala+11prb}, monolayers~\cite{fulvio+16prb}, as well as in interfaces with graphene~\cite{agg+17jpcl}.
Here we demonstrate how strongly stacking impacts the optical excitations of a vdW crystal at the absorption onset and beyond.
By varying the arrangement of individual h-BN layers, we find in total four types of electron-hole pairs, of a two-dimensional, three-dimensional (3D), and charge-transfer character, and discuss their appearance by symmetry considerations. 
We focus on the five structures that are obtained by including four inequivalent atoms in the unit cell, allowing only a rigid translation by one bond length and/or exchanged positions between the two atomic species.
We employ density-functional theory~\cite{hohe-kohn64pr,kohn-sham65pr,PBE,tkat-sche09prl,grim06jcc} and many-body perturbation theory (MBPT, including $G_0W_0$~\cite{hedi65pr,hybe-loui85prl} and the Bethe-Salpeter equation~\cite{hank-sham80prb,stri88rnc,rohl-loui00prb,pusc-ambr02prb}), implemented in the all-electron framework of the \texttt{exciting} code~\cite{gula+14jpcm,sagm-ambr09pccp,nabo+16prb}.

\begin{figure*}
\includegraphics[width=.95\textwidth]{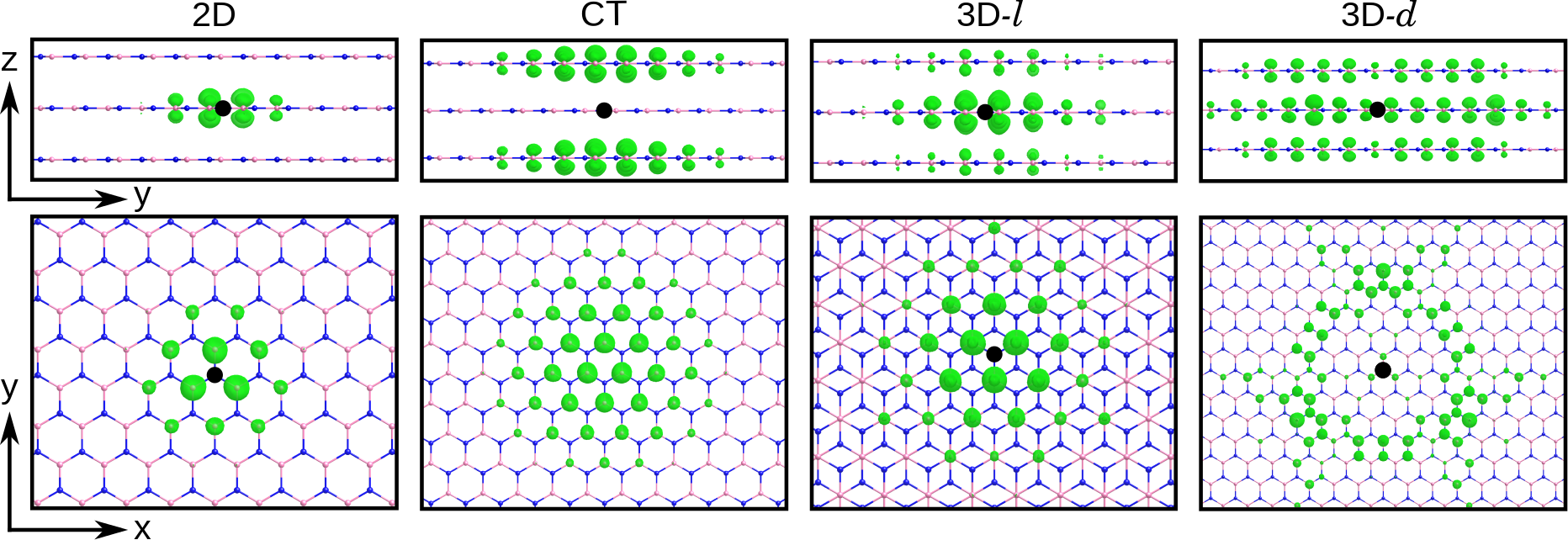}%
\caption{(Color online) Real-space distribution (side and top view) of the four types of excitons that can appear in bulk h-BN upon different layer stacking. From left to right: 2D exciton, with the electron and the hole sitting on the same layer; charge-transfer (CT) exciton, with the electron and the hole on different layers; 3D exciton with localized (3D$-l$) and delocalized (3D$-d$) in-plane extension. The hole is indicated by the black dot and the electron distribution by the green isosurface. B atoms are pink and N atoms blue. Figures produced with the VESTA software~\cite{momm-izum11jacr}.}
\label{fig:excitons}
\end{figure*}
Prototypes of the four kinds of excitations that can appear in h-BN are displayed in Fig.~\ref{fig:excitons}. 
There, the electronic distribution surrounding a fixed position of the hole is plotted. The first type is a 2D exciton, i.e., with the electron and hole on the same layer. 
Such $e$-$h$ pairs are typical of atomically-thin sheets~\cite{fulvio+16prb}, but may also dominate the absorption onset in multilayer crystals~\cite{arna+06prl,wirtz+08prl,gala+11prb,moli+15ssr,thyg17tdm}. 
We notice that these excitons are very localized, extending only up to three lattice parameters in the in-plane direction.
The trigonal shape of the excitonic wave-function reflects the hexagonal symmetry of the monolayer. 
As extensively discussed in earlier works on h-BN based on MBPT~\cite{arna+06prl,wirtz+08prl}, this exciton is twofold degenerate, and its shape results from the averaged densities of the two degenerate contributions~\cite{wirtz+08prl}.

In multilayer structures, $e$-$h$ pairs can also be extended in the vertical direction, with the electron and/or the hole spreading over neighboring layers. This is the case of \textit{charge-transfer} (CT) excitons, where the electron and the hole sit on different layers and their extension is limited to a few (here up to five) lattice parameters in the in-plane direction. 
Viewed from the top, this $e$-$h$ pair also exhibits clear trigonal symmetry (see Fig.~\ref{fig:excitons}). 
CT excitons are particularly intriguing in view of generating photocurrents~\cite{deib+10am,hong+14natn} or, if characterized by large binding energies, even Bose-Einstein condensates~\cite{min+08prb,khar+08prb,cuda+10prl,pera+13prl,fogl+14natcom}.

The other two types of excitons shown in Fig.~\ref{fig:excitons} exhibit a 3D distribution, that can be either localized in the in-plane direction (3D$-l$) or delocalized over the whole space (3D$-d$). In both cases, the electron distribution overlaps with the hole distribution and further extends uniformly and symmetrically to the adjacent layers. In the 3D$-l$ $e$-$h$ pairs, the electron probability density is significantly enhanced in the layer where the hole resides. This $e$-$h$ pair exhibits a trigonal in-plane shape with an extension similar to the one of 2D excitons. In addition to the specific arrangement of the atoms, as discussed in the following, the 3D$-l$ excitons appear in configurations where the neighboring layers are linked by inversion symmetry.
3D$-d$ excitons appear above the absorption onset and are characterized by a delocalized spatial distribution in the three directions.
In the h-BN structures considered here these $e$-$h$ pairs extend up to 12 lattice parameters in the in-plane directions and across three layers in the stacking direction~\cite{note-si}.
While reflecting a resonant character and reduced crystal symmetry, such excitons exhibit a triangular shape, typical of higher-energy excitons in monolayer h-BN~\cite{fulvio+16prb}. 

\begin{figure}
\includegraphics[width=.45\textwidth]{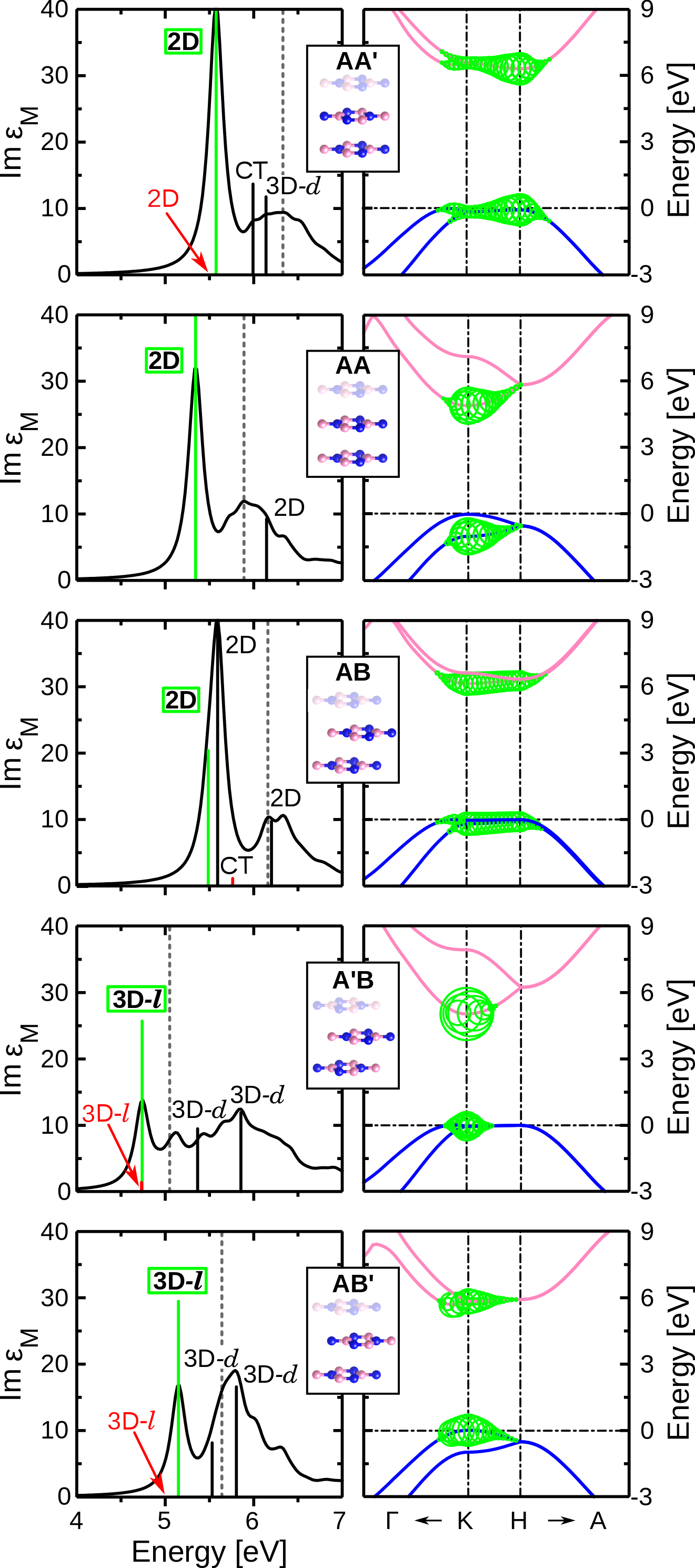}%
\caption{(Color online) Left: Optical absorption spectra, given by the imaginary part of the macroscopic dielectric function, of the five considered stacking arrangements of h-BN (insets). The character of selected excitations is indicated following the nomenclature in Fig.~\ref{fig:excitons}. The height of the vertical bars is indicative of their relative intensity~\cite{note-si}. The first bright exciton is marked by a green bar with the corresponding label framed in green. Dark exciton at lowest energy are indicated by red arrows. A Lorentzian broadening of 0.1 eV is applied to all spectra. Direct QP gaps are marked by dashed gray lines. Right: Reciprocal-space distribution of the first bright exciton with band character highlighted in color: blue for N and pink for B. The size of the green circles indicate the \textbf{k}-resolved band contributions to the corresponding excitation (green bar)~\cite{note-si}. Fermi energy set to zero at the valence-band maximum and marked by a dashed-dotted line.}
\label{fig:spectra}
\end{figure}

In order to relate the exciton characters described above with symmetry and atomic arrangement of the considered stackings, it is instructive to inspect more closely the corresponding geometries~\cite{note-si,marom+10prl,hod+12jctc} and analyze their electronic structure and optical spectra. 
In the simplest configuration, all atoms of the two inequivalent layers are aligned on top of each other (AA stacking).
In the AB or Bernal stacking, every second atoms lies on the \textit{hollow} site.
These patterns, initially defined for monoatomic stacks of graphitic layers~\cite{bern24prsla}, can be further modulated by the positions of the two atomic species. 
The AA configuration results in the AA' stacking~\cite{solozh+95ssc} if B and N atoms of alternating layers lie on top of each other. We note in passing that this is considered to be the most stable h-BN arrangement~\cite{pease+50nat,liu+03prb,Constantine+13prl}.
The A'B and AB' stackings are obtained from the AB configuration: Instead of alternating B and N atoms on the \textit{hollow} site, as in the AB pattern, in the A'B (AB') arrangement only N (B) atoms occupy the \textit{hollow} position.
While the AB configuration has been reported for bilayer structures~\cite{warner+10acsn}, other arrangements are observed only locally in few-layer stacks~\cite{kim+13nl1,shme+13ns,khan+16ns,khan+17mt,henc+17apl}.
Complex modulated patterns have been achieved in combination with graphene~\cite{kim+13nl,tang+13sr,gao+15natcom,khan+17nt,woff+17sr,zhen+17apl} while transitions from a stacking sequence to another one have been shown upon morphological deformation~\cite{bourrellier+14acs}.
These different layer stackings combined with atomic arrangements in the vertical direction strongly impact the electronic structure (Fig.~\ref{fig:spectra}, right panel) and the optical spectra accordingly.
In the following, we discuss how this structure-property relation determines the conditions under which different types of $e$-$h$ pairs are formed.
Alternative approaches to extract this information are based on model Hamiltonians~\cite{fulvio+16prb,deil-thyg18nl,pale+18condmat}, analytic envelope-function modeling~\cite{aror+17natcom}, and double-Bader analysis~\cite{wang+18jcp}.
 
In Fig.~\ref{fig:spectra} we display the optical spectra (left panel) together with the quasi-particle (QP) band structures (right panel).
Our results are in good agreement with available experimental data for the AA' stacking~\cite{mamy+81jppl,tarrio-schn89prb,solo+01jpcs,evans+08jpcm,wata-tani09prb,muse+11pssrrl,edgar+14jcg,cass+16natp,cunn+18prm}, see also Ref.~\cite{note-si}. 
First, we consider the character of the first bright exciton in each configuration (green bars). 
The contributions of individual electronic states to the $e$-$h$ pairs are indicated by the green circles drawn on top of the QP band structures~\cite{note-si}. 
Regardless of the stacking, the direct QP gap of bulk h-BN is always along or in the vicinity of the K-H path in the Brillouin zone (BZ). 
Hence, all excitations comprised within the first peak and up to $\sim$1 eV above always stem from transitions between QP states in the gap region. The two uppermost valence bands (VB, VB-1) and the two lowest conduction bands (CB, CB+1) are energetically very close to each other, due to the presence of two atoms of the same species in the unit cell. 
These bands have a well-defined N and B character (highlighted in color in Fig.~\ref{fig:spectra}), which is related to the atomic structure of the constituting species and thus independent of the layer stacking. 
 
The first bright $e$-$h$ pair has a 2D character in the AA', AA, and AB configurations. 
In the AA' stacking, it is twofold degenerate, owing to the symmetry of the lattice. 
Due to the parity of the exciton with respect to the inversion symmetry of the crystal, a doubly degenerate dark exciton with 2D distribution appears lowest in energy (here, at 5.50 eV; for a comparison with experiments see Ref.~\cite{note-si}). 
These findings are in agreement with earlier studies of this h-BN phase~\cite{wirtz+08prl,bers+13prb,pale+18condmat}. 
The 2D character of the first bright exciton is closely related to the electronic structure of the corresponding arrangements.
In the AA' and AB stackings, where atoms of the same species never lie on top of each other, VB-1 and VB (and likewise CB and CB+1) are (almost) degenerate along the entire K-H path, as reported also in Refs.~\cite{blas+95prb,liu+03prb,arna+06prl,gao12ssc}.
At the high-symmetry point H the Kohn-Sham wave-functions sit on one inequivalent layer only.
In the AB configuration the same behavior is found also at K, with the consequence that in this stacking the hole and the electron of the 2D exciton are always localized within one specific h-BN sheet.
Conversely, in the AA' and AA arrangements, electronic wave-functions tend to be delocalized over all layers towards the high-symmetry point K.
As a result, the 2D exciton appears with the same probability in any layer~\cite{note-si,note-AA}.

The lowest-energy exciton is optically allowed in those stackings that lack inversion symmetry between the layers. 
This is the case of the AB arrangement, where the absence of inversion symmetry allows for the presence of two optically-active $e$-$h$ pairs, both twofold degenerate and encompassed within the first peak (see Fig.~\ref{fig:spectra}). 
Likewise, in the AA stacking the lowest-energy exciton is optically-allowed and bears 2D distribution. 
It dominates the absorption onset giving rise to a sharp peak. 
Conversely, in the AA' stacking, exhibiting inversion symmetry, the first exciton is dark.
Overall, the spectra of these three structures look quite similar: They are characterized by an intense excitonic peak around 5.5 eV, followed by a broader and less intense hump at higher energies. 
The binding energy of the first optically-active exciton ranges between 0.50 and 0.75 eV, depending on the specific structure~\cite{note-si}.
We define the binding energy as the difference between the excitation energy computed with the $e$-$h$ interaction included (by solving the Bethe-Salpeter equation) and neglected (independent QP approximation)~\cite{note-si}.
This definition is general and holds for any excitation independent of its position in the spectrum.
For the first bright excitons discussed above, it coincides with the often adopted definition of binding energy as the difference between fundamental and optical gaps.

In the A'B and AB' configurations the lowest-energy bright exciton has a 3D$-l$ nature. 
This $e$-$h$ pair has the same in-plane extension as the 2D one, but the electron is vertically spread over two neighboring layers to the one hosting the hole. 
This distribution can be once again traced back to the symmetry and character of the electronic states near the gap that contribute to it. 
As discussed above, in the A'B (AB') stackings layers are staggered with B (N) atoms vertically aligned on top of each other. 
This makes the CB (VB) energetically split from the CB+1 (VB-1). 
The wave-functions of these bands along the K-H path are therefore delocalized within all layers~\cite{note-si}. 
Under these conditions, the resulting $e$-$h$ pairs extend to the neighboring layers.

The inversion symmetry in these structures is responsible for the presence of a forbidden exciton that is lowest in energy: In the A'B lattice, the first dark $e$-$h$ pair has the same energy as the first bright 3D$-l$ exciton, while in the AB' stacking it is found at approximately 5 eV. 
The intensity of the first peak is considerably lower compared to the other arrangements, due to the reduced wave-function overlap between the QP states contributing to it. In the A'B arrangement, the direct QP gap and thus the absorption onset are significantly red-shifted compared to the AA' stacking, up to 1.3 eV and 0.8 eV, respectively. 
Depending on the stacking arrangement, only 2D or 3D$-l$ excitons can appear at the absorption onset and are always associated with intense peaks. 
Their oscillator strength is related to the symmetry conditions as well as to the large overlap between the $\pi$/$\pi^*$ wave-functions of the bands involved in the corresponding transitions.
Their spatial localization, in turn, reflects the bound character of these $e$-$h$ pairs.
On the other hand, CT and 3D$-d$ $e$-$h$ pairs emerge only at higher energies and, especially the former, upon strict symmetry and structural requirements, reflecting the layer-selective distribution of the contributing bands. 
In fact, CT excitons can appear only in the AA' and AB stackings, since in these structures the condition that atoms of the same species are aligned on top of each other is never satisfied. 
In the AA' arrangement, the CT exciton marked in the spectrum of Fig.~\ref{fig:spectra} originates from transitions between the VB-1 and the CB+1, and between the VB and the CB, at the H point in the BZ. Here, these bands are degenerate and the corresponding electronic wave-functions are spread over N and B atoms belonging to different h-BN layers. 
The CT exciton is dipole-allowed only in the out-of-plane polarization direction and does not exhibit any degeneracy. As a result, the in-plane distribution of this $e$-$h$ pair differs from the one shown in Fig.~\ref{fig:excitons}~\cite{note-si}. In the spectrum of the AB arrangement, we find a doubly-degenerate CT exciton with in-plane polarization, stemming from transitions between the VB and the CB in the vicinity of the K point. 
According to the wave-function distribution of these bands, the hole and the electron sit on different layers, giving rise to the CT character of this exciton. 
Its weak oscillator strength is due to the small overlap between the electron and hole wave functions~\cite{note-si}. 

Delocalized 3D excitons characterize mainly the high-energy window of the optical spectra. 
They only appear when inversion symmetry is present. In the AA' configuration the 3D$-d$ exciton marked above 6 eV is twofold degenerate and has a binding energy of 0.2 eV. 
Viewed from the side (Fig.~\ref{fig:excitons}), this type of exciton extends symmetrically to the nearest-neighboring layers in the vertical direction. 
This in-plane distribution appears in the h-BN monolayer~\cite{fulvio+16prb} and is apparently preserved also in multilayer structures. Also in the A'B and AB' configurations 3D$-d$ excitons are present in the high energy-range, but they exhibit a different in-plane distribution that reflects the structural arrangement~\cite{note-si}.
It is worth noting that in the AA and AB configurations, where inversion symmetry is absent, delocalized 2D excitons with trigonal shape appear in the high energy-range.
Remarkably, they resemble higher-energy excitons in monolayer h-BN~\cite{fulvio+16prb,note-si}.

Finally, above 1 eV from the absorption onset of each stacking very delocalized excitations appear (not shown, see also Ref.~\cite{note-si}). 
They stem from a number of mixed transitions between electronic states far from the QP gap and do not correspond to any type of exciton depicted in Fig.~\ref{fig:excitons}. 
Due to their resonant character of band-to-band transitions, their delocalized distribution resembles that of Kohn-Sham states. 
Such 3D excitations with an extremely extended character are expected to form more favorably in vdW crystals and heterostructures exhibiting superlattices and/or Moir\'e patterns. 
These complex structural arrangements, that are observed in realistic samples (see, \textit{e.g.}, Refs.~\cite{yank+12natp,wood+14natp,wall+15ap} for graphene/h-BN interfaces), tend to decrease the symmetry of the system, thereby promoting weakly-bound $e$-$h$ pairs.

The general arguments underpinning the presented analysis can be directly extended to other vdW materials.
While group-IV monoatomic layers (graphene, silicene, etc.) exhibit semi-metallic character~\cite{aoki+07ssc,xu+10nt}, binary compounds such as transition-metal dichalchogenides have a pronounced excitonic behavior largely influenced by layer stacking~\cite{he+14prb,rash+14ra,rigo+15nl,yan+15nl,zhen+15prb,Zhuo+16pe,alex+17nl,mill+17nl,naya+17nano,deil-thyg18nl,gill-maul18prb,toru+18condmat}.
Spin-orbit coupling~\cite{xiao+12prl,moli+13prb,hsu+18natcom} and an indirect-direct band-gap transition upon an increasing number of layers~\cite{Mak+10prl,rupp+14nl} further enrich their excited-state properties. 
Also group-V vdW materials, such as phosphorene, are known for the dependence of their gap on the number of layers and the stacking order~\cite{Dai+14jpcl,Lei+16nl,kuma+16prb,Koda+17jpcc,lin+17jmcc,deniz+15prb,keci+16prb,mogu+16cms}.
Layer displacement in these cases can evidently further contribute to tune the character of the $e$-$h$ pairs.


In summary, we have analyzed the four types of excitations that appear in different stacking arrangements of bulk h-BN, considered here as a prototypical vdW material.
We have shown that localized $e$-$h$ pairs with a purely 2D character appear in all configurations.
Charge-transfer excitons occur only if specific structural and symmetry conditions are fulfilled. 
3D excitons with a more or less extended spatial distribution are also present at the absorption onset and/or above it, depending on the layer stacking.
Our results demonstrate the interplay between structural arrangements and optical properties in stacked vdW materials, providing the key elements to assess, foresee, and tune the character of their $e$-$h$ pairs.
As such, our findings contribute to the fascinating perspectives of designing vdW heterostructures with customized characteristics, achieved through a controlled modulation of the electronic structure via layer patterning.

Input and output files of our calculations can be downloaded free of charge from the NOMAD repository at this link: \url{http://dx.doi.org/10.17172/NOMAD/2018.06.05-1}.

\subsection*{Acknowledgment} 
This work was supported by the Algerian Ministry of High Education and Scientific Research under the PNE program. Partial funding by the German Research Foundation (DFG), through the Collaborative Research Center 951, HIOS, is appreciated. W.~A.~thanks F. Paleari for fruitful discussions.
 

\newpage
\newpage
\section{SUPPLEMENTAL MATERIAL:
Dimensionality of excitons in stacked van der Waals materials: The example of hexagonal boron nitride}

\author{Wahib Aggoune}
\affiliation{Institut f\"{u}r Physik and IRIS Adlershof, Humboldt-Universit\"{a}t zu Berlin, Berlin, Germany}
\affiliation{Laboratoire de Physique Th\'eorique, Facult\'e des Sciences Exactes, Universit\'e de Bejaia, 06000 Bejaia, Algeria}
 
\author{Caterina Cocchi}
\author{Dmitrii Nabok}
\affiliation{Institut f\"{u}r Physik and IRIS Adlershof, Humboldt-Universit\"{a}t zu Berlin, Berlin, Germany}
\affiliation{European Theoretical Spectroscopic Facility (ETSF)}

\author{Karim Rezouali}
\author{Mohamed Akli Belkhir}
\affiliation{Laboratoire de Physique Th\'eorique, Facult\'e des Sciences Exactes, Universit\'e de Bejaia, 06000 Bejaia, Algeria}

\author{Claudia Draxl}
\affiliation{Institut f\"{u}r Physik and IRIS Adlershof, Humboldt-Universit\"{a}t zu Berlin, Berlin, Germany}
\affiliation{European Theoretical Spectroscopic Facility (ETSF)}
\maketitle

%
%
\section*{Theoretical Methods and Computational Details}
Ground-state properties are computed in the framework of density functional theory (DFT), within the generalized gradient approximation for the exchange-correlation functional with the Perdew-Burke-Ernzerhof parameterization \cite{PBE}.
The Tkatchenko-Scheffler (vdW-TS) approach~\cite{tkat-sche09prl} is adopted to account for van der Waals interactions between layers.
The DFT-D2 method by S. Grimme~\cite{grim06jcc} is used to estimate binding energies. 
Optical spectra are obtained in the framework of many-body perturbation theory.
Quasi-particle (QP) energies are computed within the $G_0W_0$ approximation \cite{hedi65pr,hybe-loui85prl}. 
Optical spectra are obtained from the solution of the Bethe-Salpeter equation (BSE), an effective two-body equation for the electron-hole two-particle Green's function \cite{hank-sham80prb,stri88rnc}.
The BSE Hamiltonian reads $H^{BSE} = H^{diag} + 2H^{x} + H^{dir}$, where the first term $H^{diag}$ accounts for \textit{vertical} transitions, while the other two terms incorporate electron-hole \textit{exchange} ($H^{x}$) and the screened Coulomb interaction ($H^{dir}$). 
The excitation energies $E^{\lambda}$ are the eigenvalues of the secular equation associated to the BSE Hamiltonian
\begin{equation}
\sum_{v'c'\mathbf{k'}} H^{BSE}_{vc\mathbf{k},v'c'\mathbf{k'}}A^{\lambda}_{v'c'\mathbf{k'}} = E^{\lambda}A^{\lambda}_{vc\mathbf{k},}
\label{eq:ham}
\end{equation}
where $v$ and $c$ indicate valence and conduction states, respectively. 
The eigenvectors $A^{\lambda}$ enter the expression of the transition coefficient of each excitation $\lambda$:
\begin{equation}
\mathbf{t}_{\lambda} = \sum_{vc\mathbf{k}} A^{\lambda}_{v c \mathbf{k}} \frac{\langle v \mathbf{k} \vert \widehat{\mathbf{p}} \vert c \mathbf{k} \rangle}{\epsilon_{c \mathbf{k}}\ -\ \epsilon_{v \mathbf{k}}}.
\label{eq:osci} 
\end{equation}
This quantity carries information about the polarization of the excitation in each direction.
The oscillator strength of a given excitation corresponds to the absolute square of $\mathbf{t}_{\lambda}$. 
This quantity is represented in Fig.~\ref{fig:str1}c together with all spectra. 
On the other hand, in Fig. 2 (main text) and in Figs.~\ref{fig:exc1},~\ref{fig:exc2}, the height of the bars corresponds to $|\mathbf{t}_{\lambda}|$.
This representation is adopted to ease the display of the oscillator strength of the selected excitations within the spectra. 
$\mathbf{t}_{\lambda}$ also enters the expression of the imaginary part of the macroscopic dielectric function,
\begin{equation}
\mathrm{Im}~\varepsilon_M~=~\dfrac{8\pi^2}{\Omega} \sum_{\lambda} |\mathbf{t}_{\lambda}|^2 \delta(\omega - E^{\lambda}),
\label{eq:abs}
\end{equation}
where $\Omega$ is the unit cell volume.
The BSE eigenvectors $A^{\lambda}$ also contain information about the $\mathbf{k}$-resolved contributions of individual QP bands to the electron-hole pairs, expressed by the coefficients 
\begin{align}
& w^{\lambda}_{v\mathbf{k}} = \sum_c |A^{\lambda}_{vc\mathbf{k}}|^2, \\
& w^{\lambda}_{c\mathbf{k}} = \sum_v |A^{\lambda}_{vc\mathbf{k}}|^2.
\label{eq:weight}
\end{align}
for valence and conduction bands, respectively.
These coefficients, also called here \textit{exciton weights}, are visualized by green circles in the band-structure plots of Fig. 2, main text, and in Figs.~\ref{fig:exc1},~\ref{fig:exc2} below.
Their size is representative of the contribution of a specific transition to the corresponding exciton.
BSE eigenvectors $A^{\lambda}$ also provide information about character and spatial extension of excitons in real space, acting as coefficients in the expression of the two-particle excitonic wave-functions: 
\begin{equation}
\Psi^{\lambda}(\mathbf{r}_{e},\mathbf{r}_{h}) = \sum_{v c \mathbf{k}} A^{\lambda}_{vc\mathbf{k}}\phi_{c\mathbf{k}}(\mathbf{r}_{e})\phi^{*}_{v\mathbf{k}}(\mathbf{r}_{h}).
\label{eq:psi}
\end{equation}
Different from the Kohn-Sham states that retain the crystal periodicity, excitonic wave-functions are non-periodic.
Their finite extension needs to be properly resolved numerically through the $\mathbf{k}$-point sampling.

\begin{figure*}[hbt]
\includegraphics[width=.85\textwidth]{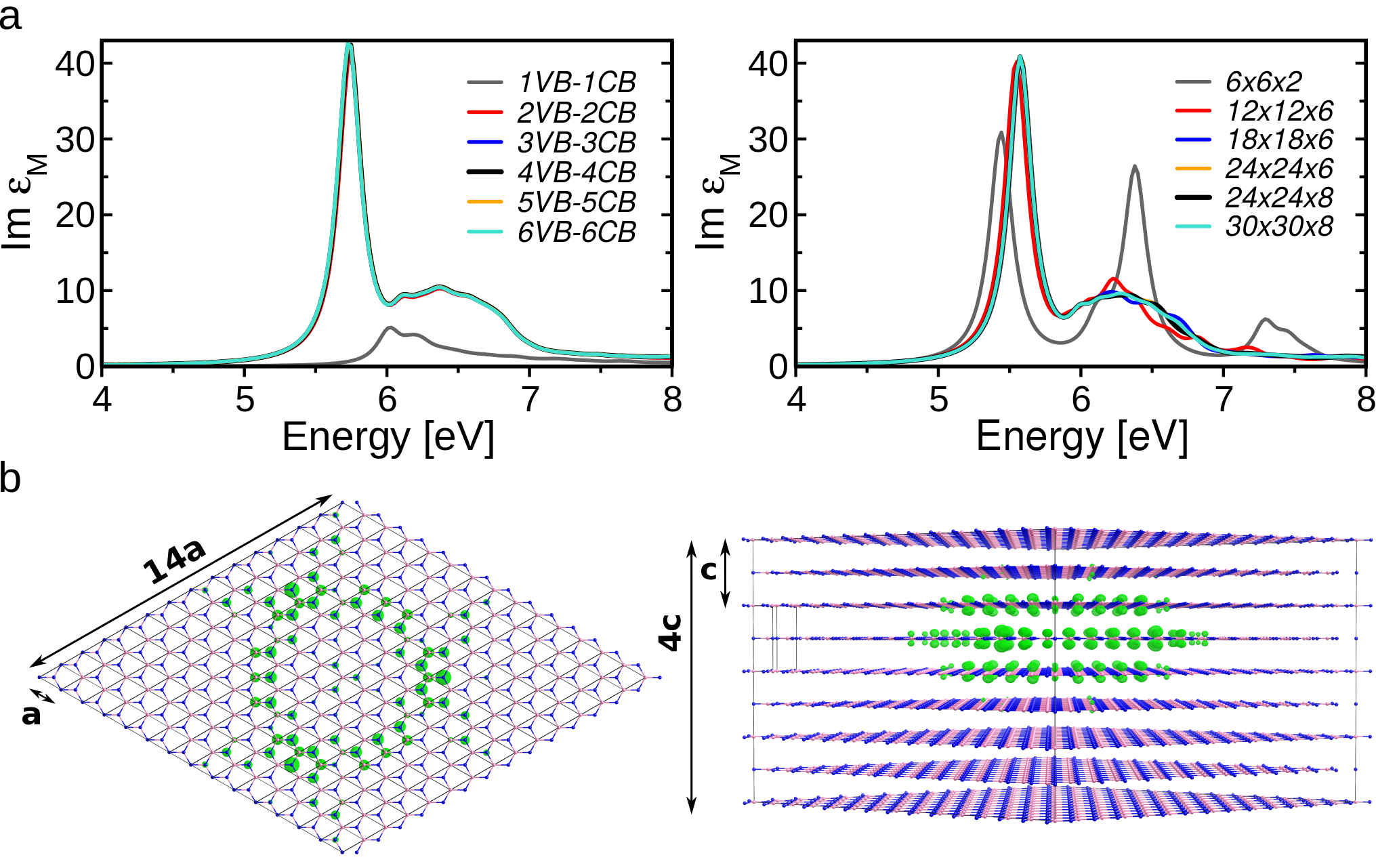}%
\caption{\textbf{a}: BSE convergence for the optical spectrum of h-BN, AA' stacking. Left: Convergence with respect to the number of valence and conduction bands (VB and CB, respectively) obtained on a 18$\times$18$\times$6 \textbf{k}-mesh shifted from the $\Gamma$-point. These calculations are performed on top of the Kohn-Sham band-structure by applying a scissors operator. Right: Convergence with respect to the size of the \textbf{k}-mesh considering transitions from 4 occupied to 4 unoccupied bands. These calculations are performed on top of the QP band structure. \textbf{b}: Real-space distribution of the 3D-d exciton of the AA' stacking shown in Fig. 1 in the main text. Left: Top view, with the in-plane unit cell parameter $a$ and supercell parameter $14a$ highlighted. Right: Side view, with the out-of-plane unit cell parameter $c$ and supercell lattice parameter $4c$ highlighted. The electron distribution for a fixed hole position is given by the green isosurface.}
\label{fig:convergence}
\end{figure*}

All calculations are performed using {\exciting} \cite{gula+14jpcm}, an all-electron full-potential code, implementing the family of linearized augmented planewave plus local orbitals methods.
In the ground-state calculations, a basis-set cutoff R$_{MT}$G$_{max}$=7 is used.
For both atomic species involved, namely boron (B) and nitrogen (N), a muffin-tin radius R$_{MT}$=1.3 bohr is adopted.
The sampling of the Brillouin zone (BZ) is performed with a 36 $\times$ 36 $\times$ 14 $\textbf{k}$-grid. 
Lattice constants and internal coordinates are optimized until the residual forces on each atom are smaller than 0.003 eV/\AA{}. 
Calculations of the QP correction to the Kohn-Sham eigenvalues within the $G_{0}W_{0}$ approximation \cite{nabo+16prb} are performed with 250 empty states and a BZ sampling with a 18 $\times$ 18 $\times$ 6  shifted $\textbf{k}$-mesh is adopted.
These parameters ensure a numerical accuracy over the QP gap at the high-symmetry point H of about 20 meV.
For the solution of the BSE \cite{sagm-ambr09pccp} a plane-wave cutoff  R$_{MT}$G$_{max}$=6 is employed.
In the calculation of the response function and of the screened Coulomb potential 100 empty bands are included. 
In the construction and diagonalization of the BSE Hamiltonian 4 occupied and 4 unoccupied bands are considered and a 24 $\times$ 24 $\times$ 8 shifted $\textbf{k}$-point mesh is adopted (see Fig.~\ref{fig:convergence}a below for some convergence tests).
This choice ensures well converged spectra and an accuracy on the energy of the first exciton of at least 10 meV.
Local-field effects are taken into account by including 25 $|\mathbf{G}+\mathbf{q}|$ vectors.
A Lorentzian broadening of 0.1 eV is applied to the resulting spectra.
Atomic structures and isosurfaces are visualized with the VESTA software \cite{momm-izum11jacr}.
For the real-space visualization of the electron-hole pairs (Eq.~\ref{eq:psi}), a supercell with in-plane lattice parameter of 14$a$ and out-of-plane lattice parameter of 4$c$ has been used, with $a$ and $c$ being the lattice vectors of the corresponding unit cell (see Fig.~\ref{fig:convergence}b).
In this way, eight layers are always included in the stacking direction for visualizing the excitons.
In all such plots presented here and in the main text, the position of the hole is fixed according to the composition of the exciton, and the correlated electron distribution is visualized by the isosurface (green).
The adopted $\mathbf{k}$-point sampling and the related supercell size are sufficient to ensure a meaningful representation of the types of e-h pairs depicted in Fig. 1 of the main text and in Figs.~\ref{fig:exc1} and~\ref{fig:exc2} below.
For representing the real-space distribution of all excitons we adopt isolvalues of 12$\%$ of the maximum values.

\section*{Structural and Electronic Properties}
\begin{table*}[hbt]
\centering 
 \begin{tabular}{c|ccccccccccccccccccc}
\hline \hline
\multirow{1}{*}{h-BN bulk}               &\quad\quad&\quad\quad AA'&\quad\quad AA&\quad\quad AB&\quad\quad A'B&\quad\quad AB' \\
\hline
\multirow{2}{*}{Lattice parameter (\AA)}& \quad $a$ &\quad\quad 2.50   &\quad\quad 2.50     &\quad\quad 2.50     &\quad\quad 2.50    &\quad\quad 2.50      \\
                                    & \quad $c$  &\quad\quad 6.60 &\quad\quad 6.84     &\quad\quad 6.58     &\quad\quad  6.48  &\quad\quad 6.86      \\

\multirow{1}{*}{Inversion symmetry}& \quad $i$ &\quad\quad $\surd$   &\quad\quad X     &\quad\quad X     &\quad\quad  $\surd$    &\quad\quad $\surd$      \\
\hline
\multirow{1}{*}{Binding energy (eV)}& \quad $E_{b}$ &\quad\quad 0.355   &\quad\quad 0.260 &\quad\quad 0.360     &\quad\quad  0.340  &\quad\quad  0.270     \\
\hline \hline
\end{tabular}
\caption{Lattice parameters and binding energies of bulk h-BN in the considered stacking arrangements}
\label{tab1}
\end{table*}

In Table~\ref{tab1}, we summarize stability and structural parameters of bulk h-BN, in different stacking arrangements. 
The AA stacking can be characterized by only one layer periodically repeated in the out-of-plane direction. 
Here, we consider on purpose for this system a unit cell with two layers in order to study the effects of stacking consistently with all other arrangements. 
The in-plane lattice parameter $a$=2.50 \AA{} remains constant for all structures, while the out-of-plane one, $c$, strongly depends on the layer stacking. 
The structural parameter of the AA' stacking (given by vdW-TS method) are in agreement with the experimental one reported in Ref.~\cite{solozh+95ssc}.

\begin{figure*}[hbt]
\includegraphics[width=.8\textwidth]{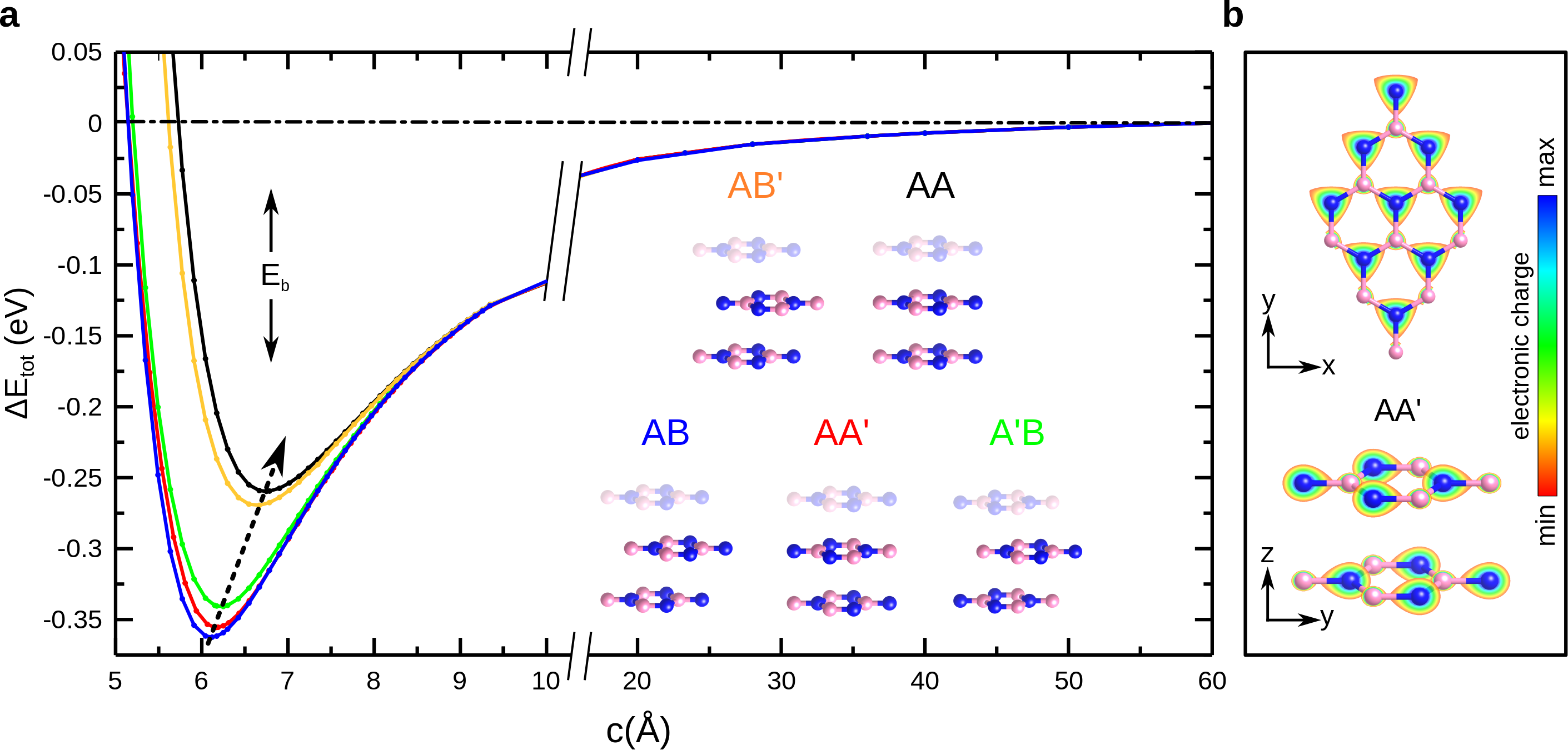}%
\caption{\textbf{a}: Binding energies of the five considered h-BN stackings (structures shown as insets) evaluated as the total energy per unit cell at different interlayer distances along the $c$ direction. The total energy of $c=60$ \AA{} is set to zero for reference. Different stackings are identified by different colors. \textbf{b}: In- and out-of-plane projection of the electronic charge density in AA' arrangement.}
\label{fig:stability}
\end{figure*}

\begin{figure*}[h!]
\includegraphics[width=.92\textwidth]{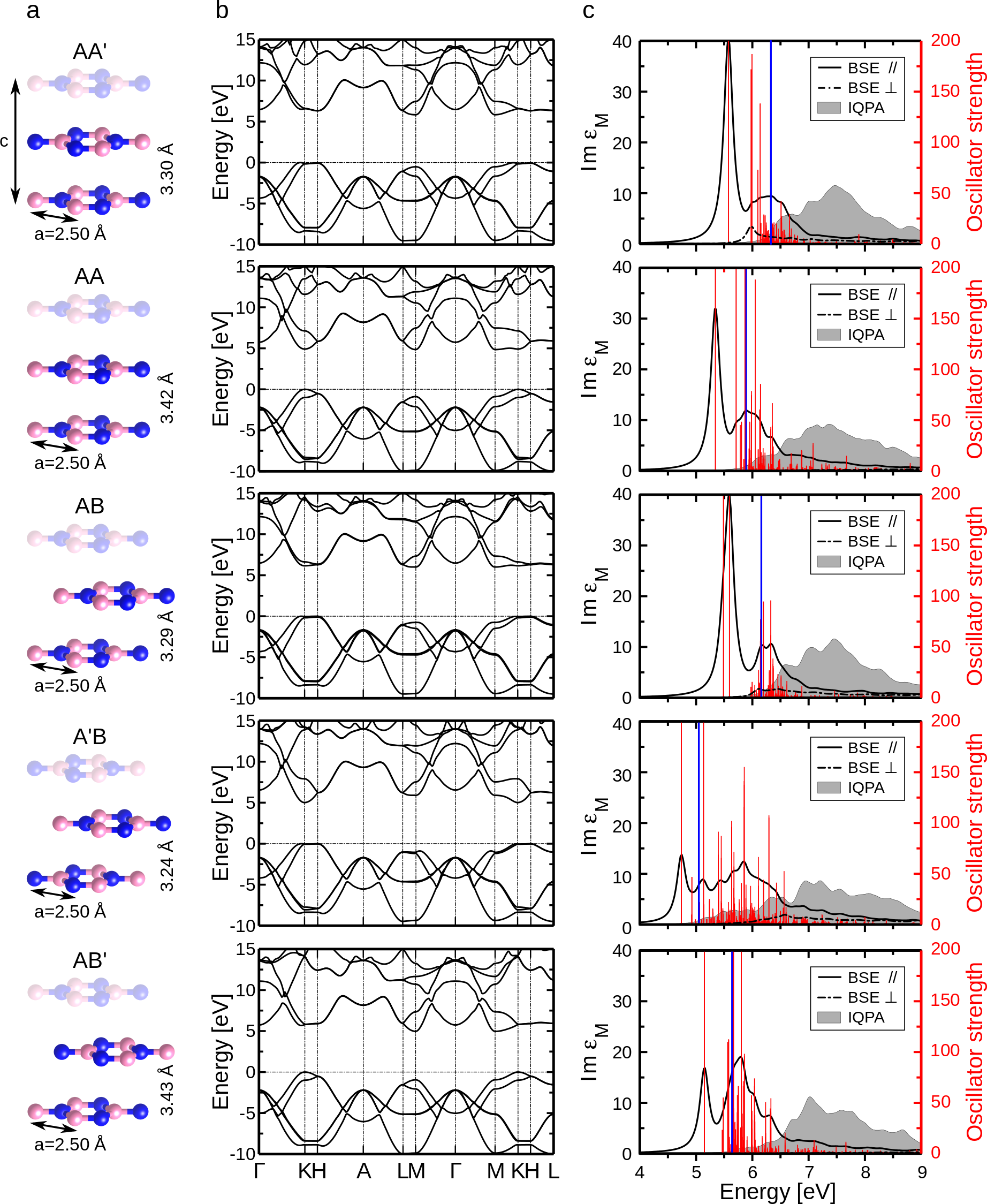}%
\caption{\textbf{a}: Sketch of the unit cells of the five considered stacking arrangements of bulk h-BN. The two layers included in the unit cell are shown in the foreground, while periodic images are shaded. B atoms are pink, N are blue. \textbf{b}: Quasi-particle band structures with valence-band maximum set to zero. \textbf{c}: Optical spectra computed by including excitonic effects (BSE: solid line for the in-plane component and dotted-dashed line for the out-of-plane one) and neglecting them within the independent QP approximation (IQPA, shaded area). Red bars indicate the full oscillator strength of the BSE solutions  (see Eq. \ref{eq:osci}) and the blue line the direct QP gap.}
\label{fig:str1}
\end{figure*}

Due to the enhanced electronegativity compared to the B atoms, there is an excess of charge in the vicinity of the N atoms, enhancing the polarity of the B-N bond (see Fig.~\ref{fig:stability}b).
As a result, in addition to the vdW forces which anchor the layers at fixed distance~\citep{marom+10prl}, the electrostatic interactions between the vertically aligned atomic centers play an important role in the interlayer distance of different stackings~\citep{hod+12jctc}. 
For this reason, the AA (with N over N and B over B) and AB' (with N over N) stackings exhibit a larger $c$ lattice parameter compared to the other configurations, due to the large electrostatic repulsion (see Fig.~\ref{fig:str1}).
We notice that the Pauli repulsion between the more delocalized electron clouds of the N atoms further contributes to increase the interlayer distance~\cite{hod+12jctc}, which is also reflected in the smaller binding energies (trend indicated by the dashed arrow in Fig.~\ref{fig:stability}a).
The most unstable configuration is indeed the AA stacking, followed by the AB' one. 
In the AA', AB, and A'B arrangements, where the N atoms are not aligned on top of each other, electrostatic repulsion forces are weaker giving rise to the largest binding energies. Moreover, since N and B atoms are vertically aligned in the AA' and AB arrangements, the electrostatic attraction between the oppositely charged atomic centers make them the most favorable configurations, as found in Ref.~\cite{Constantine+13prl, liu+03prb}, with binding energy up to $\sim$0.36 eV (see Table \ref{tab1} and Fig. \ref{fig:stability}). 
We notice that, since bulk h-BN crystallizes in the AA' configuration~\cite{pease+50nat,Constantine+13prl,liu+03prb}, the larger binding energy found in the AB arrangement with respect to the AA' stacking is due to the accuracy limits of our method~\cite{marom+10prl,Constantine+13prl,hod+12jctc}. 
The AB arrangement has been reported to appear in bilayer h-BN~\cite{warner+10acsn}. 
Recently, even the other stackings considered in this work have been probed experimentally in few-layer h-BN structures~\cite{kim+13nl1,shme+13ns,bourrellier+14acs,khan+16ns,khan+17mt,henc+17apl}. 

A more detailed investigation of structural properties and stability of these systems goes beyond the scope of this work and will be addressed in a dedicated study.

\begin{table*}[hbt]
\centering
 \begin{tabular}{c|ccccccccccccccccccc}
\hline \hline
\multirow{1}{*}{h-BN bulk}               &\quad\quad&\quad\quad AA'&\quad\quad AA&\quad\quad AB&\quad\quad A'B&\quad\quad AB' \\
\hline
\multirow{2}{*}{Electronic gap (eV)}& \quad DFT &\quad\quad  4.22  &\quad\quad 3.30     &\quad\quad 4.35     &\quad\quad  3.48  &\quad\quad 3.43      \\
                                    & \quad \textit{$G_{0}W_{0}$}  &\quad\quad   5.83    &\quad\quad 4.84 &\quad\quad 6.00     &\quad\quad  5.03     &\quad\quad 4.96 \\
\hline
\multirow{2}{*}{Minimum direct gap (eV)}   & \quad DFT &\quad\quad   4.63 &\quad\quad 4.16 &\quad\quad 4.46    &\quad\quad  3.50 &\quad\quad 4.00 & &    \\
                                    & \quad \textit{$G_{0}W_{0}$}  &\quad\quad  6.33  &\quad\quad 5.89 &\quad\quad 6.16    &\quad\quad  5.05 &\quad\quad 5.64 & &    \\   

\hline
\multirow{2}{*}{Excitation energy (eV)}   & \quad 1$^{st}$ exciton ($\times$2) &\quad\quad   \textbf{5.50} &\quad\quad 5.34 &\quad\quad 5.48    &\quad\quad  \textbf{4.73} &\quad\quad \textbf{5.04} & &    \\
                                    & \quad 2$^{st}$ exciton ($\times$2) &\quad\quad  5.57  &\quad\quad 5.71 &\quad\quad 5.59    &\quad\quad  4.74 &\quad\quad 5.14 & &    \\                      
                                                
\hline  \hline
\end{tabular}
\caption{Electronic and direct QP gaps of h-BN bulk in the five considered stacking arrangements as well as the excitation energy of the first and the second exciton, with ($\times$2) to indicate that they are doubly degenerate. Dark excitons are in bold.}
\label{tab2}
\end{table*}

In the first column of Fig.~\ref{fig:str1} we show a sketch of the unit cells of the five considered h-BN stackings and in second column their corresponding QP band structures.
While the QP band structures of different stackings look quite similar to each other, we notice two important differences. 
The first one concerns the character and the size of the band gap.
As reported in Table~\ref{tab2}, depending on the stacking the difference between the electronic and the direct (optical) gap can be more or less pronounced.
An indirect band gap of 5.83 eV between the valence-band maximum (VBM) near the K point and the conduction-band minimum (CBM) at the M point is found in the AA' stacking of h-BN.
Our result is in agreement with the previous theoretical findings based on MBPT~\cite{blas+95prb,arna+06prl,gao12ssc,pale+18condmat}. 
While the reported experimental band gaps range between 3.6 and 7.1 eV~\cite{solo+01jpcs}, the size and the indirect nature of the band gap is in agreement with recent experimental observations~\cite{edgar+14jcg,evans+08jpcm,cass+16natp}. 
We also notice that the reported band gaps of AA' and other stackings are in good agreement with the $GW$ results in Refs.~\cite{gao12ssc,pale+18condmat}.
Also, comparing the gaps of the different structures we notice a relevant decrease in those stackings where atoms of the same species are vertically aligned.
This is the case of the A'B and AB' configurations, as well as of the AA one.
It is worth noting here that the latter structure is formed by only one inequivalent h-BN layer.
As mentioned in the main text, we simulate this system including two layers in the unit cell to treat this arrangement consistently with all the others. 
As a result, the first optically allowed transition is the one between the second uppermost valence band (VB-1) and the lowest conduction band (CB).
In Table~\ref{tab2} we report the value of 5.89 eV (4.16 eV), corresponding of the minimum direct QP (DFT) gap between the VB-1 and the CB at the high-symmetry point K, consistent with that given by the IQPA spectrum, as shown in the following. This value is in agreement with Ref.~\cite{bourrellier+14acs}.

The second difference in the electronic properties of the considered arrangements is related to the energy separation between VB-1 and VB, as well as CB and CB+1 along the K-H path. 
Since all the unit cells contain 2 B and 2 N atoms, the N-like bands (VB-1/VB) and the B-like bands (CB/CB+1) are split (degenerate) along the K-H path when the atoms of the same species in unit cell are (not) on top of each other. 
This energy splitting reflects the weak electrostatic interaction between atoms of the same species.

\begin{figure*}
\includegraphics[width=.95\textwidth]{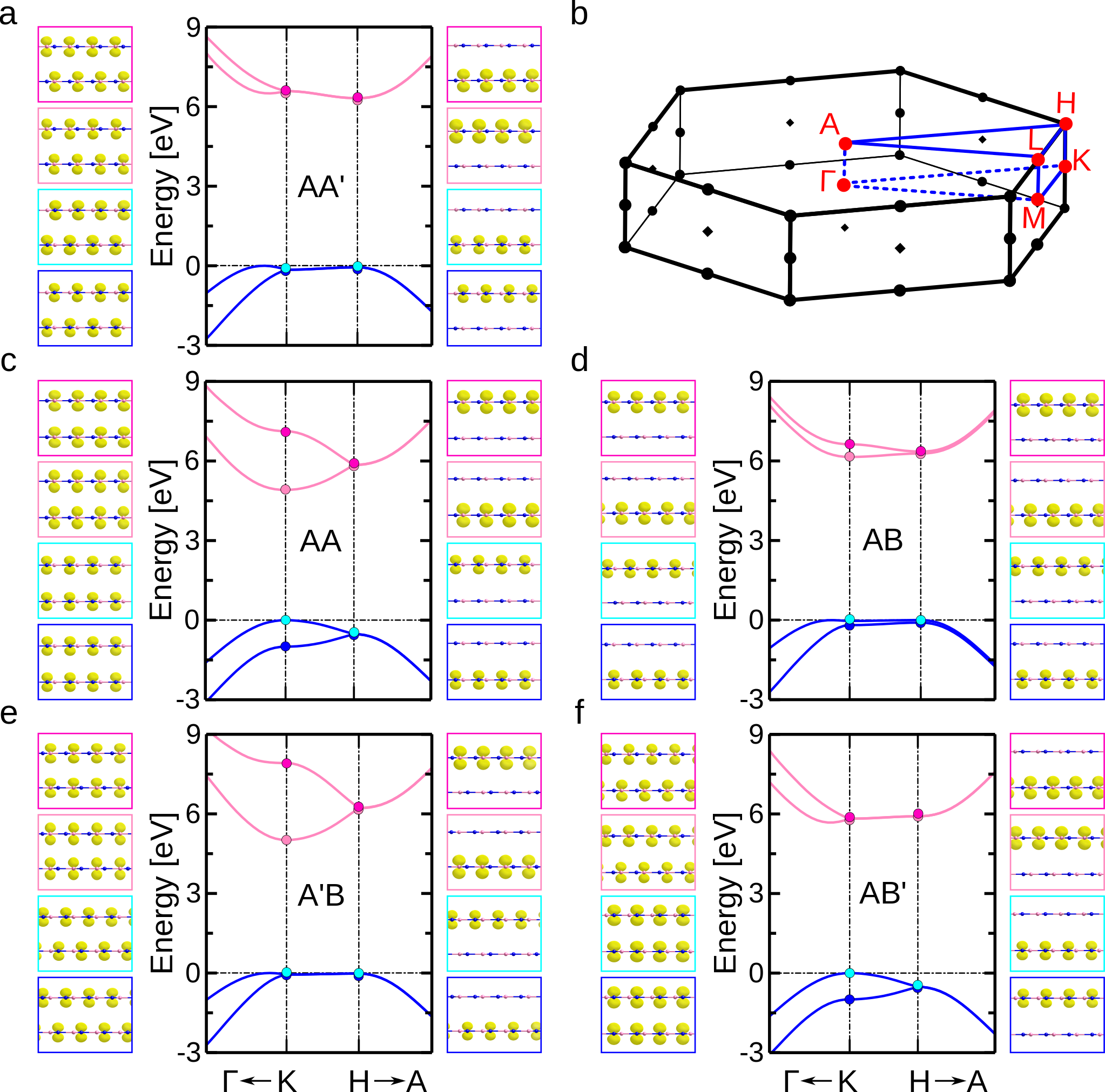}%
\caption{(a,c,d,e,f) QP band structures in the vicinity of K-H path of the five considered stacking arrangements and real-space distribution of the two highest occupied and the two lowest unoccupied bands at the K (left) and at the H (right) point in the BZ. The band character is indicated by the color code of the corresponding atomic species (B: pink, N: blue), while the box frame points to the corresponding electronic states marked in the band structures. The valence-band maximum is set to zero. (b) Sketch of the Brillouin zone with the high-symmetry points highlighted.}
\label{fig:KS}
\end{figure*}

To visualize the effect of layer arrangement on the band structure of h-BN, we plot the real-space distribution of the VB-1/VB and CB/CB+1 at the K and H points (see Fig.~\ref{fig:KS}). 
At the H point the VB-1/VB and the CB/CB+1, with N- and B-like character, respectively, are degenerate in all stackings. 
The corresponding electronic distribution of each band is spread over only one h-BN layer in unit cell. 
At the K point, the VB-1/VB and the CB/CB+1 are either energetically degenerate or split, depending on the stacking. 
The electronic distribution of these bands is spread uniformly over both h-BN layers in the unit cell of all stackings, except for the AB arrangement. 
Here, the VB-1/VB and the CB/CB+1 are split at the K point and are therefore spread over only one h-BN layer in unit cell, like at the H point.

\section*{Optical Properties}

The differences in the structural and electronic properties between the stackings are reflected in the optical spectra.
The in-plane and out-of-plane component of the imaginary part of the macroscopic dielectric function are shown for each system in the third column of Fig.~\ref{fig:str1}. 
We notice that in all stackings a sharp excitonic peak appears in the in-plane spectra, followed by a broad hump formed by inter-band transitions, and thus visible also in the independent QP spectra.
 In the AA', A'B and AB' arrangements, which exhibit inversion symmetry, the first peak is given by the second twofold degenerate bright exciton (see Table~\ref{tab2}).
The first exciton is also twofold degenerate and optically-forbidden due to its parity with respect to the inversion symmetry operation. 
Due to the lack of inversion symmetry in the AB stacking, the first peak in the corresponding spectrum embraces degenerate lowest-energy bright excitons. 
Also in AA arrangement, the first exciton is twofold degenerate optically allowed and gives rise to the intense excitonic peak dominating the absorption onset. 
While the intensities of the low-energy peak in the AA, AA', and AB stackings are similar, their energies differ due to the varying direct QP gap (see Table~\ref{tab2}).
In the A'B and AB' stackings the spectrum is red-shifted compared to the other structures. Concomitantly, the lower symmetry of these two configurations and the reduced wave-function overlap between the contributing QP states are responsible for the weaker peak intensities and for the larger number of allowed excitations. 
The corresponding solutions of the BSE (red bars in Fig.~\ref{fig:str1}) are mainly dipole-active but have weak intensity, in contrast with the AA, AA', and AB stackings, where most excitations are either very intense or forbidden.
The out-of plane component of the imaginary part of the macroscopic dielectric function is given by a dotted-dashed line (Fig.~\ref{fig:str1}). 
While these spectra are rather featureless in the relevant energy-region for this study, in the AA' stacking a relatively intense peak appears around 6 eV.
It corresponds to the non-degenerate charge-transfer exciton discussed in the main text.

\begin{figure*}
\includegraphics[width=.95\textwidth]{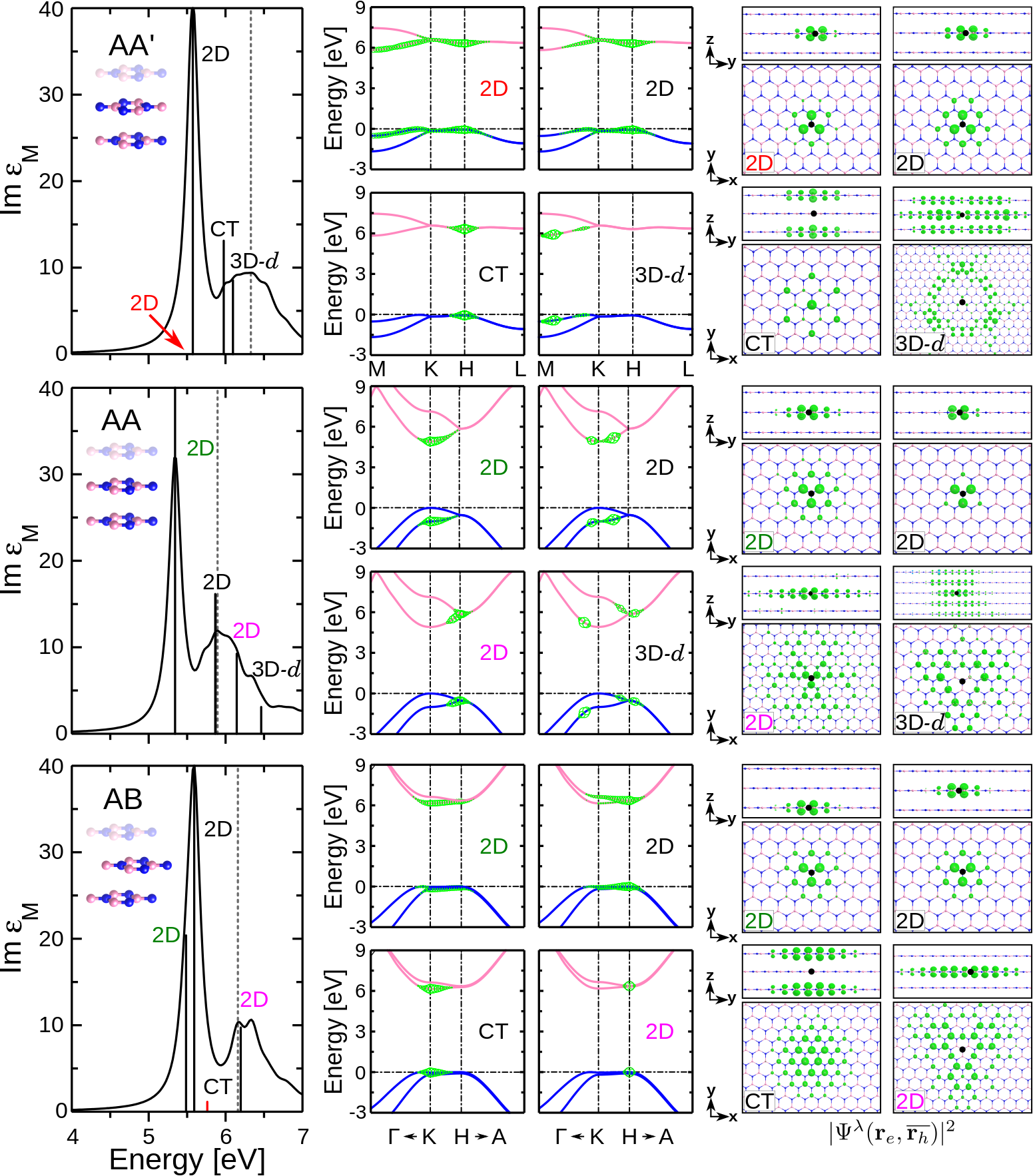}%
\caption{(Left): Optical spectra of the AA', AA, and AB stacking of h-BN (structures in the inset), computed by including excitonic effects (BSE). A Lorentzian broadening of 100 meV is included. The dotted line indicates the direct QP gap and the intensity of the black bars indicate the relative oscillator strength given by $|t_{\lambda}|$ (see Eq. \ref{eq:osci}). (Center): Band contributions to the excitations  marked in the spectra: The size of the green circles is representative of the weights (Eq.~\ref{eq:weight}) of the involved QP states. The band character is indicated by the color code of the atomic species (B: pink, N: blue). The valence-band maximum is set to zero. (Right): Three-dimensional projections of the electron component of the e-h wave-functions highlighted in the spectrum, with the position of the fixed hole marked by the black dot.}
\label{fig:exc1}
\end{figure*}

In Fig. \ref{fig:exc1} we report the absorption spectra of the AA', AA and AB stackings together with the corresponding real- and \textbf{k}-space distribution of the main excitons. 
As discussed in main body of the article, most excitations are double-degenerate for symmetry reasons. 
As such, the correlated probability density associated to the electron, with the hole being fixed, is a linear combination of the single contributions of degenerate BSE solutions.
With the excitonic spatial distribution given by the square modulus of the electron-hole wave-functions (see Eq.~\ref{eq:psi}), here, we report the corresponding averaged densities, consistent with the results reported in Ref.~\cite{wirtz+08prl}. 
The twofold degenerate 2D exciton that gives rise to the sharp peak in the AA' stacking, comes from transitions between the QP states along the K-H path in the BZ. 
In Fig.~\ref{fig:exc1} we show the \textbf{k}-space distribution of this exciton along the M-K-H-L path. In addition to the K-H path, the most relevant contribution comes from the M-K path in the BZ, consistent with the results on h-BN monolayer published in Ref.~\cite{fulvio+16prb}. 
The CT exciton in the AA' stacking is dipole-allowed only in the perpendicular polarization direction and does not exhibit any degeneracy.
The corresponding in-layer distribution of this exciton differs from the one shown in the main text, which is twofold degenerate and referred to the AB stacking. 
In the AA arrangement, in addition to the 2D exciton that dominates the absorption onset, about $\sim$1 eV above the absorption onset we find a delocalized 2D exciton with trigonal symmetry as in monolayer h-BN~\cite{fulvio+16prb}. 
The electronic distribution of the bands contributing to it (VB-1 and CB, around the H point) prevents the vertical extension of this exciton, giving rise to its 2D delocalized character. 
In the AA' stacking it has instead 3D$-d$ distribution.
At higher energy  ($\sim$1.2 eV above the absorption onset), we find a very delocalized 3D exciton (see Fig.~\ref{fig:exc1}). It arises form mixed transitions between QP states outside the gap. Due to their resonant character, their distribution resembles delocalized Kohn-Sham states, which can appears in all stackings at higher energy range.
As discussed above, in these stackings (AA' and AA), the 2D excitons that dominates the absorption onset, can appear uniformly in both h-BN layer in the unit cell since the electronic distribution of the contributing bands along the K-H path is spread over all layers (see Fig. \ref{fig:KS}). 
Differently, in AB arrangement, the excitons are localized on one specific layer in the unit cell according to the wave function distribution of the contributing bands along the K-H path (see Fig. \ref{fig:KS}). 
The 2D exciton that forms the sharp peak and the lowest-energy one are localized on different h-BN layers, as confirmed by the electronic wave-function distribution shown in Fig. \ref{fig:KS}.
They come, respectively, form VB $\rightarrow$ CB+1 and VB-1 $\rightarrow$ CB transitions along the K-H path in the BZ. 
The CT exciton, characterized by a weak oscillator strength, stems from transitions between the VB and the CB, which are spread over N and B atoms, respectively, belonging to different layers in the unit cell (see Fig. \ref{fig:KS}). 
At higher energies, above 6 eV, a weakly delocalized e-h pair appears with 2D character. 
In contrast to the AA' stacking that exhibits inversion symmetry and where thus the 3D$-d$ is extended to the neighboring layers, here 
the in-plane nature of this exciton is determined by the electronic distribution of the bands contributing to it: They preserve the trigonal symmetry reported also in the monolayer~\cite{fulvio+16prb}. 
At the higher energies ($\sim$1.2 eV above the absorption onset) excitons with very delocalized 3D distribution appear, resembling the delocalized Kohn-Sham states appearing in this stacking. 
They stem from mixed transitions between QP states outside the gap region.

\begin{figure*}
\includegraphics[width=.95\textwidth]{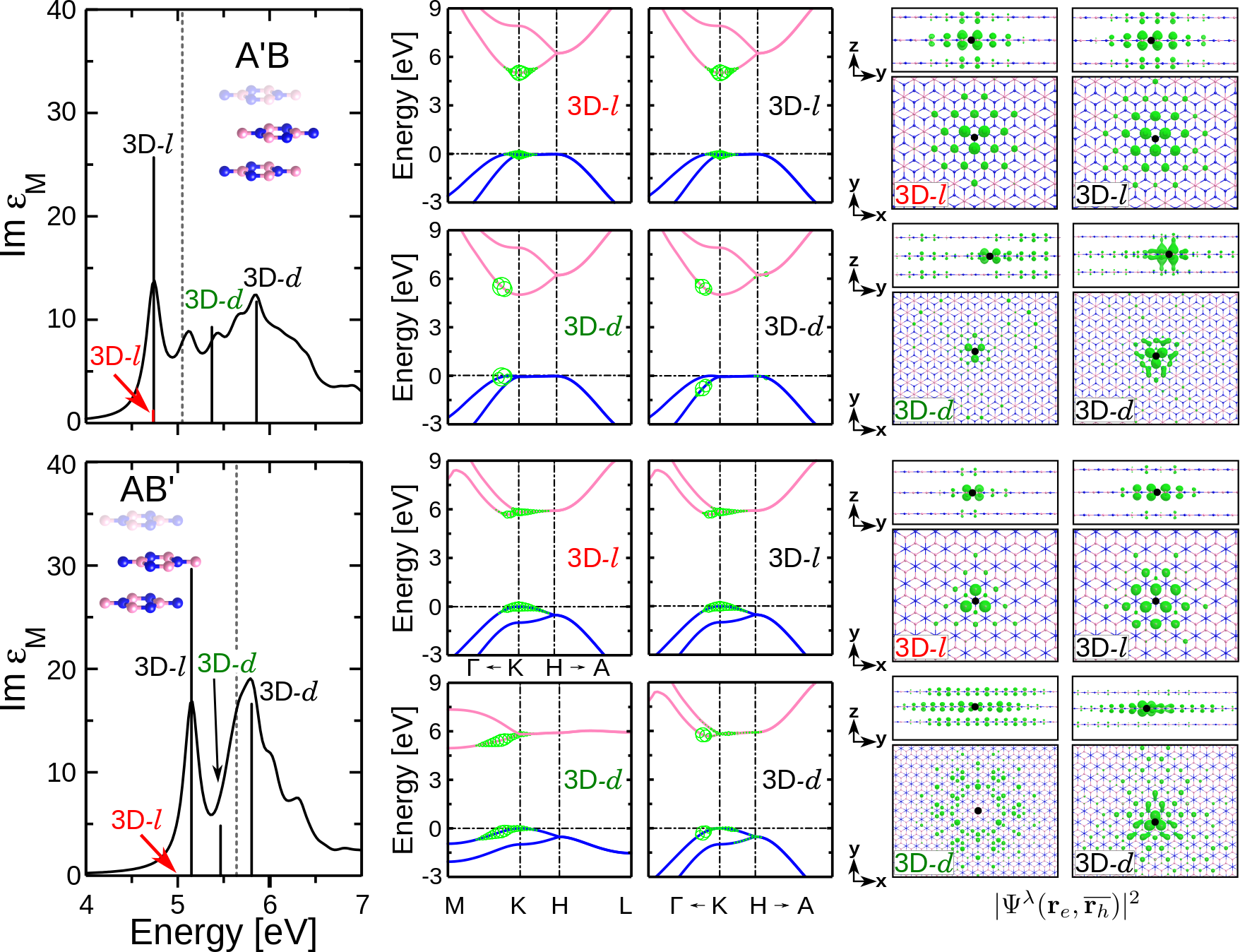}%
\caption{(Left): Optical spectra of the A'B and AB' stackings of h-BN (structures shown in the inset), computed by including excitonic effects (BSE). A Lorentzian broadening of 100 meV is included. The dotted line indicates the direct QP gap and the intensity of the black bars indicate the relative oscillator strength given by $|t_{\lambda}|$ (see Eq. \ref{eq:osci}). (Center): Band contributions to the excitations marked in the spectra: The size of the green circles is representative of the weights (Eq.~\ref{eq:weight}) of the involved QP states. The band character is indicated by the color code of the atomic species (B: pink, N: blue). The valence-band maximum is set to zero. (Right): Three-dimensional projections of the electron component of the e-h wave-functions highlighted in the spectrum, with the position of the fixed hole marked by the black dot.}
\label{fig:exc2}
\end{figure*}

In Fig. \ref{fig:exc2} the same analysis is presented for the A'B and AB' configurations.
Since the electronic distribution of the QP states along the K-H path is spread over both staggered h-BN layers in the unit cell (the layers are linked by inversion symmetry) and due to the reduced wave-function overlap between the QP states, these stackings exhibit a large number of weak excitations, mostly characterized by a 3D-like distribution.
In the spectra of these systems the absorption onset is dominated by a 3D$-l$ exciton which exhibits an in-layer distribution analogous to the one of the 2D e-h pair in the AA' stacking. 
Above the sharp excitonic peak, a 3D$-d$ exciton appears. 
Its in-plane distribution preserves the trigonal symmetry, similar to higher-energy excitons in monolayer h-BN~\cite{fulvio+16prb}.
Furthermore, it extends uniformly and symmetrically over the nearest neighboring layers, as in AA' stacking. 
This exciton arises form transitions between the VB and the CB outside the K-H path in the BZ, where the bands are split and their corresponding wave-functions are spread over both h-BN layers in the unit cell.

\section*{Comparison with experiments}

\begin{figure*}
\includegraphics[width=.95\textwidth]{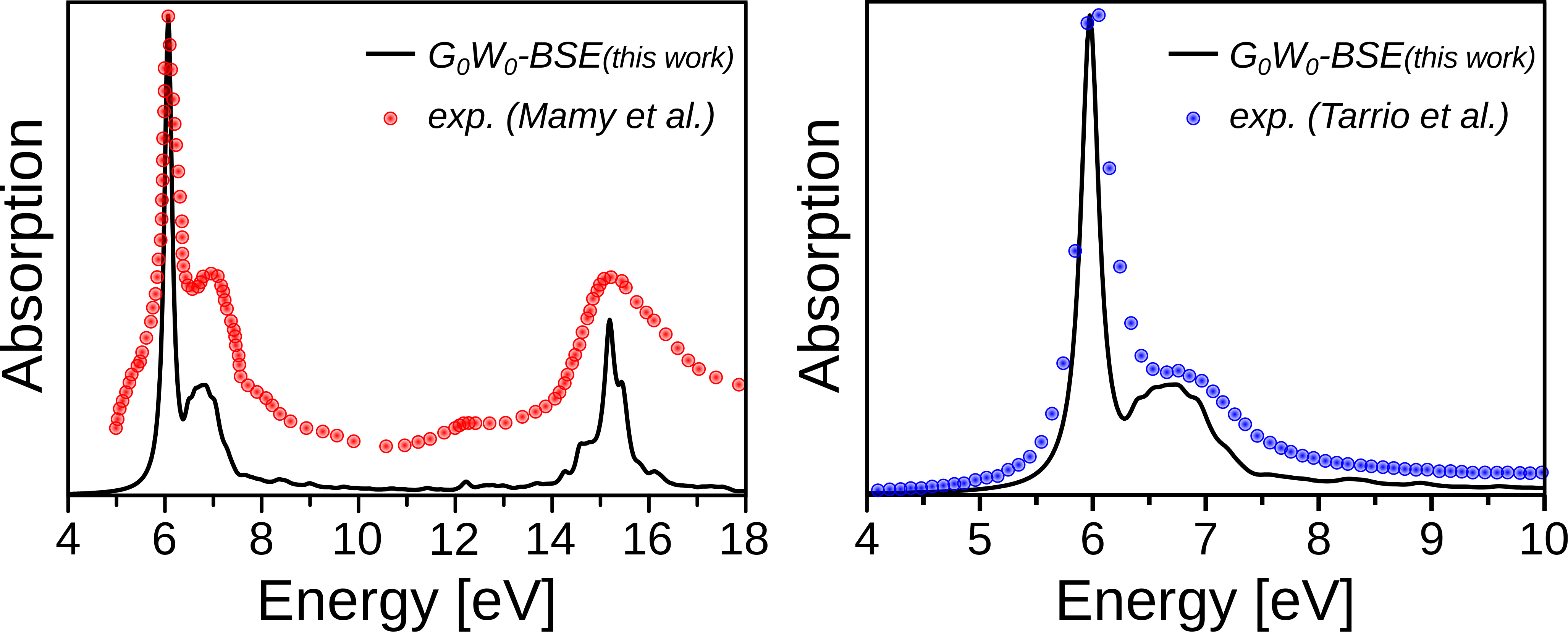}%
\caption{ Calculated optical spectrum of bulk h-BN in the AA' stacking compared with experimental data from (left) Mamy \textit{et al.}~\cite{mamy+81jppl} and (right) Tarrio and Schnatterly~\cite{tarrio-schn89prb}. The computed spectrum is blue-shifted by 0.495 eV (left) and 0.4 eV (right) to align with the measured position of the first peak. The spectral intensity is aligned to the height of the first peak.}
\label{fig:BSE-vs-exp}
\end{figure*}

For a final assessment of the quality of our results we briefly discuss them in the context of the experimental literature. 
Most of the available measurements refer to the most stable AA' stacking configuration of h-BN, which we consider for this analysis.
In Fig.~\ref{fig:BSE-vs-exp} we plot our spectra in comparison with two experimental datasets obtained with different techniques and spanning different energy regions.
The intensity of the first bright peak is aligned to match the measured values.
In both cases our computed spectrum is in agreement with the measurements.
All spectral features are accurately resolved, also the relative intensity of each maximum and minimum.
In Fig.~\ref{fig:BSE-vs-exp}, right panel, the region around the first sharp peak and the subsequent hump is plotted. 
Experimental data from Ref.~\cite{tarrio-schn89prb} are obtained with inelastic X-ray scattering at finite $\mathbf{q}$.
The spectrum shown here is taken at $\mathbf{q}$=0.13 \AA{}$^{-1}$ but its similarity to the one at $\mathbf{q}$=0, discussed in the original reference~\cite{tarrio-schn89prb} makes this comparison meaningful~\cite{cunn+18prm}.
The experimental spectrum plotted in Fig.~\ref{fig:BSE-vs-exp}, left panel, is obtained by optical absorption measurements and covers a much wider energy range, including also the second pronounced maximum at about 15 eV~\cite{mamy+81jppl}.
All features are correctly resolved in our calculation, including the weak peak at about 12 eV.
As a side remark, our spectra are blue-shifted by 0.4 -- 0.5 eV to match the experimental onset of both datasets in Fig.~\ref{fig:BSE-vs-exp}.
In addition to the uncertainty of $\pm$0.2 eV in the experimental spectrum from Ref.~\cite{tarrio-schn89prb}, we ascribe this difference to the intrinsic underestimation of the gap given by the single-shot $G_{0}W_{0}$ method compared to a self-consistent treatment~\cite{bers+13prb,cunn+18prm}.
The binding energy of the first bright exciton of about 0.75 eV (see Table~\ref{tab2}) further contributes to shift the peak position to lower energy compared to experiment~\cite{cunn+18prm}.
We rule out any artifact produced by the computational parameters adopted in the $G_{0}W_{0}$ calculations (see also Ref.~\cite{pale+18condmat}).
On top of this, the nature of the band-gap in bulk h-BN is still under debate.
Our $G_{0}W_{0}$ results for the AA' stacking indicate an indirect band gap of 5.83 eV between the VBM near the K point and the CBM at the M point and a direct QP gap of 6.33 eV in the vicinity of M.
As mentioned above, this result is in the range of experimental measurements~\cite{evans+08jpcm,muse+11pssrrl,edgar+14jcg,cass+16natp,wata-tani09prb} and in agreement with a recent $G_{0}W_{0}$ study~\cite{gao12ssc,pale+18condmat}.


\begin{thebibliography}{119}
\bibitem[{not({\natexlab{b}})}]{note-AA}
\bibinfo{note}{For consistency with the other structures, also the AA stacking
  is represented in a unit cell containing two layers. On account of the
  resulting band folding, the lowest-energy transition between VB and CB is
  optically forbidden and the first allowed transition is between VB-1 and CB.}
\end{thebibliography}
\end{document}